\documentclass[journal]{IEEEtran}
 
\usepackage{babel}

\usepackage[LGR,T1]{fontenc}
\usepackage[latin9]{inputenc}
\usepackage{amsmath}
\usepackage{amssymb}
\usepackage{graphicx}

\makeatletter

\newcommand*\LyXZeroWidthSpace{\hspace{0pt}}
\DeclareRobustCommand{\greektext}{%
  \fontencoding{LGR}\selectfont\def\encodingdefault{LGR}}
\DeclareRobustCommand{\textgreek}[1]{\leavevmode{\greektext #1}}
\ProvideTextCommand{\~}{LGR}[1]{\char126#1}

\usepackage[ruled,vlined]{algorithm2e}
\usepackage{amsmath,amsfonts,amssymb}

\usepackage{caption}
\usepackage{amsmath,amsfonts}
\usepackage{array}
\usepackage{enumitem}
\usepackage{txfonts}
\usepackage{verbatim}
\usepackage{balance}
\usepackage{mathptmx}
\usepackage{textcomp}
\usepackage[math]{blindtext}
\usepackage{babel}
\usepackage{varioref}
\usepackage{newtxtt}
\usepackage{graphicx}
\usepackage{url}
\usepackage[numbers,square, comma, sort&compress]{natbib}
\usepackage{ltxnew}
\usepackage{pdfcolmk}
\usepackage{stfloats}
\usepackage{algpseudocode,caption}
\usepackage{subcaption}
\usepackage{listings}
\usepackage{import}
\usepackage[ruled]{algorithm2e}

\usepackage{caption}
\usepackage{hyphenat}
\usepackage{url}

\def\BibTeX{{\rm B\kern-.05em{\sc i\kern-.025em b}\kern-.08em T\kern-.1667em\lower.7ex\hbox{E}\kern-.125emX}}

\makeatother

\begin{document}
\title{Power Optimization in Satellite Communication Using Multi-Intelligent
Reflecting Surfaces}

  \author{Muhammad Ihsan Khalil\thanks{The author is with the School of Engineering, RMIT University, Melbourne, Australia. Email: \texttt{muhammad.khalil@rmit.edu.au}.}}

\maketitle
\begin{abstract}
This study introduces two innovative methodologies aimed at augmenting
energy efficiency in satellite-to-ground communication systems through
the integration of multiple Reflective Intelligent Surfaces (RISs).
The primary objective of these methodologies is to optimize overall
energy efficiency under two distinct scenarios. In the first scenario,
denoted as Ideal Environment (IE), we enhance energy efficiency by
decomposing the problem into two sub-optimal tasks. The initial task
concentrates on maximizing power reception by precisely adjusting
the phase shift of each RIS element, followed by the implementation
of Selective Diversity to identify the RIS element delivering maximal
power. The second task entails minimizing power consumption, formulated
as a binary linear programming problem, and addressed using the Binary
Particle Swarm Optimization (BPSO) technique. The IE scenario presupposes
an environment where signals propagate without any path loss, serving
as a foundational benchmark for theoretical evaluations that elucidate
the system's optimal capabilities. Conversely, the second scenario,
termed Non-Ideal Environment (NIE), is designed for situations where
signal transmission is subject to path loss. Within this framework,
the Adam algorithm is utilized to optimize energy efficiency. This
non-ideal setting provides a pragmatic assessment of the system's
capabilities under conventional operational conditions. Both scenarios
emphasize the potential energy savings achievable by the satellite-RIS
system. Empirical simulations further corroborate the robustness and
effectiveness of our approach, highlighting its potential to enhance
energy efficiency in satellite-to-ground communication systems.
\end{abstract}

\section{Introduction }

Satellite communication systems have been a cornerstone of global
telecommunications for several decades, providing essential services
such as broadcasting, navigation, and remote sensing. However, the
inherent challenges of satellite communication, such as long propagation
delays, signal attenuation, and interference, have motivated continuous
research and development efforts to enhance system performance and
reliability. Reconfigurable Intelligent Surfaces (RISs) have recently
garnered significant attention as a potential cornerstone for the
next generation of wireless communication networks \cite{Basharat2021}.
Essentially, an RIS is a synthetic surface laden with a vast array
of cost-effective, reconfigurable passive reflecting elements. These
elements have the capability to modify the trajectory of incoming
wireless signals by fine-tuning their amplitude and phase shift \cite{DiRenzo2020}.
One of the standout features of RISs is their independence from radio
frequency (RF) chains. This independence translates to a substantial
decrease in energy usage and hardware expenses, positioning RISs as
a more cost-effective and eco-friendly alternative to traditional
multi-antenna and relaying systems \cite{Basar2019,Khalil2022}. Given
these attributes, RISs are emerging as a streamlined and cost-efficient
solution for achieving wireless communication that boasts high spectral
and energy efficiencies. This potential has ignited a surge of interest
from both the commercial and academic sectors, all eager to harness
the full spectrum of benefits that RISs offer.

\subsubsection*{Related Works}

\paragraph{Research Insights into Single RIS-Aided Terrestrial Systems}

The domain of single RIS-aided systems has been thoroughly explored
in various studies \cite{Wu2019a,Liang2019,Basar2019,Yang2020a,Yang2020,Yang2020b,Yang2020c,Ferreira2020}.
A comparative analysis in \cite{Wu2019a} demonstrates that RIS outperforms
conventional massive multiple-input multiple-output systems and multi-antenna
amplify-and-forward (AF) relaying networks, offering advantages in
system complexity and cost reduction. Reference \cite{Liang2019}
provides insight into the basic characteristics of RIS/antenna technology
and explores its potential applications. In \cite{Basar2019}, the
authors present a comprehensive overview of state-of-the-art solutions,
highlight fundamental distinctions between RIS and other technologies,
and discuss pivotal open research issues in this field. The utilization
of RIS to enhance the quality of source signals transmitted to a destination
via an unmanned aerial vehicle is explored in {[}\cite{Yang2020a}.
Study \cite{Yang2020} investigates the performance of an RIS-assisted
mixed indoor visible light communication/radio frequency (RF) system,
deriving closed-form expressions for Outage Probability (OP) and bit
error rate (BER) for both AF and decode-and-forward (DF) relaying
schemes. In \cite{Yang2020b}, the secrecy OP of an RIS-assisted network
is derived, considering the presence of a direct link and an eavesdropper.
Conversely, \cite{Yang2020c} derives accurate approximations for
channel distributions and performance metrics of RIS-assisted networks,
assuming Rayleigh fading channels and applicability to any number
of reflecting elements. More recently, \cite{Ferreira2020} provided
closed-form expressions for the bit error probability of RIS-assisted
networks over Nakagami-m fading channels. They noted that their results
are primarily valid for BPSK and QAM modulation techniques. While
the authors in \cite{Ferreira2020} considered Nakagami-m fading channels,
their exact expressions for error probability were confined to a limited
number of reflecting elements. 

\paragraph{Multiple RISs-Aided terrestrial station}

Diverging from the exploration of single RIS-aided networks, several
scholarly works have delved into analyses involving multiple RISs.
Pertinent literature in this domain can predominantly be classified
into two main categories: those focusing on optimization and those
centered on performance analysis. In the optimization category, \cite{Mei2021}
introduces a novel approach, exploiting the line-of-sight (LoS) link
between adjacent RISs to establish a multi-hop cascaded LoS link between
the base station (BS) and the user. In this context, a set of RISs
are strategically selected to sequentially reflect the BS\textquoteright s
signal, thereby maximizing the received signal power at the user.
However, it's crucial to note that this study exclusively considered
the impact of path-loss, omitting the potential impact of fading.
Conversely, a more recent contribution to this category is \cite{Yang2022a},
which explored the optimization of RIS-aided networks by deploying
multiple RISs to cater to wireless users. The researchers sought to
enhance the network's energy efficiency by dynamically adjusting the
operational status of each RIS and fine-tuning the reflection coefficients
matrix. For further exploration into optimization challenges within
RIS-aided networks that utilize multiple RISs, \cite{Alexandropoulos2020}
and \cite{Lyu2020} provide additional perspectives and findings.

Meanwhile, the utilization of multiple RISs can enhance communication
systems by providing numerous paths for received signals, thereby
amplifying the received signal strength. The outage probability in
systems assisted by multiple RISs was scrutinized and optimized in
\cite{KumarHindustani2023}. In \cite{Wang2020}, the authors explored
the use of multiple RISs to maximize the received power for downlink
point-to-point millimeter-wave communications. Considering multi-hop
transmission through RIS, the design of double-hop assisted wireless
communication was investigated in \cite{Zheng2021} and \cite{Huang2021}.
Furthermore, the multi-cell network with multiple RISs was examined
in \cite{Ni2021}, taking into account non-orthogonal multiple access.
By concurrently considering uplink and downlink, the weighted sum
rate maximization problem for multi-RIS-assisted full-duplex systems
was studied in \cite{Saeidi2021}. However, the aforementioned works
\cite{KumarHindustani2023}-\cite{Saeidi2021}  presupposed that all
RISs are operational, which is not energy-efficient since RISs also
consume energy for signal controlling. 

\paragraph{RISs-Aided Satellites Systems}

Effective communication between satellites and terrestrial stations
is paramount for the transmission of data across extensive distances.
With the proliferation of satellites and an escalation in data transmission
requirements, it is imperative to augment the energy efficiency of
these communication systems. Satellites, predominantly powered by
photovoltaic cells and energy storage units, necessitate meticulous
energy management to prolong their operational lifespan. In an epoch
where extended satellite operability and sustainability are of paramount
importance, the enhancement of energy efficiency is indispensable.

A pragmatic approach to optimize energy utilization in satellite systems
involves the refinement of the ground segment of the communication
architecture. By augmenting signal processing capabilities at terrestrial
stations, it becomes feasible to decode attenuated signals emanating
from satellites. Consequently, satellites can operate at reduced power
levels, thereby conserving energy while concurrently ensuring robust
and reliable communication. This methodology underpins the sustainable
functionality of satellite systems and guarantees the integrity of
data transmission, thereby upholding the efficacy of global communication
networks. In recent advancements, the integration of Reflecting Intelligent
Surfaces (RISs) into satellite communication frameworks has been proposed,
offering an auxiliary indirect path for signal transmission concomitant
with the direct Line-of-Sight (LoS) link from satellites. This integration
has the potential to fortify the received signal and mitigate the
effects of environmental perturbations \cite{Bariah2022}. While studies
like \cite{Aldababsa2023,Niu2023,Lin2022} and \cite{Najafi2021}
have highlighted the benefits of RISs in terrestrial wireless communication,
there is limited academic research on their use in satellite communication
systems. The unique characteristics of satellite communications, encompassing
global coverage, protracted propagation distances, and the spherical
geometry of Earth, present novel challenges and opportunities for
the implementation of RISs \cite{Kodheli2021a}. The scarcity of research
specifically addressing the integration of RISs in satellite systems
underscores a significant gap in the current body of knowledge, emphasizing
the need for further exploration in this innovative domain.

\subsection*{Contribution}

Energy Efficiency ($\eta$) is defined as the ratio of power received
by the system to its total power consumption \cite{Whitaker2005}.
While the energy efficiency of RIS-aided satellites has been explored
to some extent, comprehensive research in this domain remains sparse
in the extant literature. This investigation endeavors to address
this lacuna by proposing a new model tailored to augment the energy
efficiency of multi-RIS-aided satellite communication systems. Several
determinants influence $\eta$ in RIS-aided satellites, encompassing
the power expended per reflecting element, RIS phase shifting consumption,
control power operation, and the energy requisitioned by the associated
circuitry. It is salient to note that the predominant power consumption
in an RIS emanates from its reflecting elements. As delineated by
\cite{Tang2021}, each RIS reflection element necessitates approximately
0.33 mW of power. Consequently, the incorporation of multiple RISs
invariably escalates consumption due to the augmented count of reflecting
elements. In light of this backdrop, the present manuscript elucidates
an optimization framework dedicated to amplifying the system's energy
efficiency. This is achieved through the deployment of two distinct
methodologies under varied environmental paradigms, thereby enriching
the academic discourse with diverse optimization techniques.

\subsection*{First Method: Ideal Environment (IE)}

In the Idle Environment (IE) method, we propose an optimization scheme
that presupposes an environment devoid of perturbations. Here, \textquotedbl IE\textquotedbl{}
denotes a scenario wherein the signal, emanating from both the satellite
and the RIS panel and directed towards the user, remains unimpeded
by path-loss, shadowing, and other factors that could impact signal
propagation. The optimization of Energy Efficiency ($\eta$) in this
context is delineated into two sub-problems: the maximization of power
received by the end-user and the minimization of the overall power
expenditure. This bifurcation facilitates a more granular analysis,
thereby enabling the formulation of targeted solutions that strike
an equilibrium between performance and $\eta$. To optimize power
reception in the IE method, we introduce a methodology to calculate
the received power, accounting for both the Line-of-Sight (LoS) and
the RIS-reflected components. Subsequently, we employ this initial
computation to ascertain the optimal phase shifts for each RIS, with
the aim of achieving maximal power reception. Utilizing Selective
Diversity (SD), we identify the RIS configuration that yields the
highest power reception. Conversely, the minimization of power consumption,
predominantly attributed to the reflecting elements, is formulated
as a binary linear programming problem. In addressing this challenge,
we employ the Binary Particle Swarm Optimization (BPSO) algorithm,
acclaimed for its efficacy in navigating complex solution spaces to
identify optimal configurations. Upon the concurrent maximization
of power reception and minimization of power consumption, we deduce
the optimal energy efficiency.

\subsection*{Second Method: Non-Ideal Environment (NIE)}

Within the Non-Ideal Environment (NIE) framework, we consider both
deterministic and stochastic components of signal path loss. Deterministic
path loss predominantly arises from signal attenuation, influenced
by factors such as transmission distance and medium-specific properties,
including air or vacuum. Conversely, stochastic path loss is attributed
to shadowing and various unpredictable environmental dynamics. For
the optimization of $\eta$ in the NIE context, we utilize the Adam
algorithm, a methodology originally conceptualized for neural networks
by \cite{Kingma2017}. In its foundational application, the Adam algorithm
measures the divergence between predicted and actual outputs, typically
denoted as a loss, serving as the objective function. The algorithm
capitalizes on moving averages of the parameters, termed momentums,
to expedite optimization. Moreover, its adaptive learning rates, which
adjust based on historical gradient data, ensure meticulous convergence
towards the optimal solution.

In the context of our proposed system, the primary objective is to
enhance $\eta$. In this endeavor, the attributes of the Adam algorithm,
specifically its adaptive learning rates and momentum components,
prove invaluable. These features facilitate the accurate calibration
of optimal parameter values, potentially offering a more efficient
approach than conventional optimization methods. Considering the complexity
of the proposed system, which comprises multiple RISs, each with an
array of reflection elements, the proficiency of the Adam algorithm
in navigating complex optimization challenges is manifest. Given these
considerations, we have adapted the Adam algorithm into a unique strategy
designed to estimate energy efficiency in RIS-aided satellite networks,
especially under NIE conditions. The outcomes from this approach validate
the algorithm's proficiency in RIS-centric scenarios, and our simulation
results provide additional validation of its reliability.

The structure of this paper is as follows: Section \ref{sec:Syst_Mod}
presents the system model, elaborating on the components and configuration
of the RIS-assisted communication system. Section \ref{sec:EE } articulates
our proposed formulation for energy efficiency, encompassing both
the received power and power consumption metrics. Section \ref{ideal },
emphasizes the optimization of the RIS phase shift under ideal conditions,
elucidating the methodologies and techniques employed to optimize
energy efficiency. Section \ref{sec: EE_non idel} explores the intricacies
of non-ideal scenarios, with a particular focus on the implications
of path-loss and environmental uncertainties on received power. The
paper concludes with Section \ref{Conclusion}, which synthesizes
our findings and proposes potential directions for future research
and advancements in the domain of RIS-assisted wireless communication
systems.

\section{System Model\label{sec:Syst_Mod} }

\begin{figure}
\centering{}\includegraphics[viewport=3bp 0bp 700bp 350bp,width=3.6in]{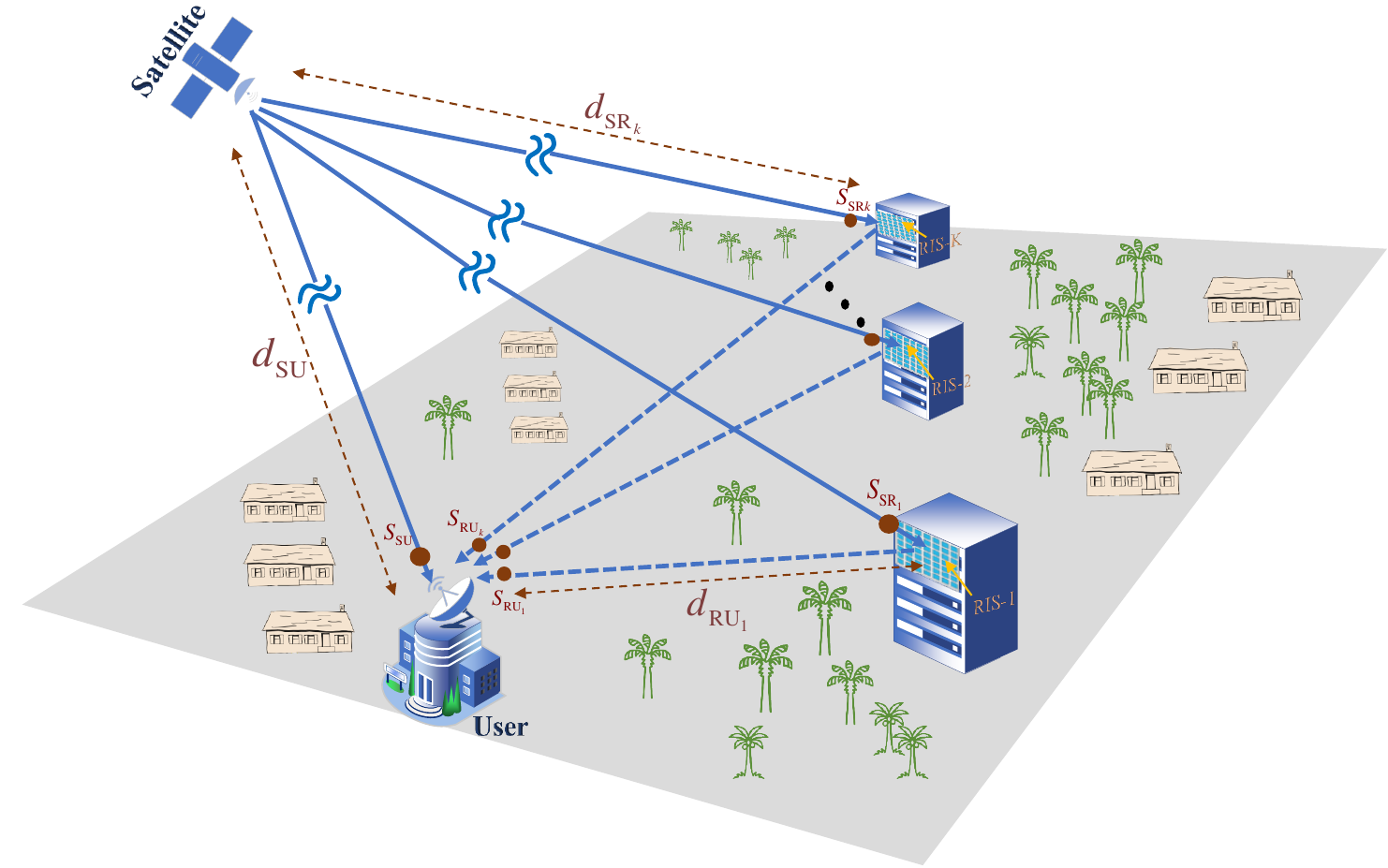}\caption{Geometry model for the proposed Multiple-RIS- assisted satellite\label{fig: 0}}
\end{figure}

In this study, we present a satellite communication system designed
to transmit data to a terrestrial user. This system incorporates multiple
Reflecting Intelligent Surfaces (RISs) situated on the Earth's surface
to facilitate the relay of the satellite signal to the user. communications.
Each RIS is composed of $N$ elements, which can be dynamically adjusted
to modulate both the phase and amplitude of the reflected electromagnetic
waves, thereby creating an indirect transmission route that complements
the\textbf{ }LOS link arriving directly from satellite communications.
As illustrated in Fig. \eqref{fig: 0}, the system's configuration
adopts a triangular arrangement that includes the Satellite (S), the
User receiver (U), and a designated terrestrial RIS. Within this arrangement,
the signal propagation paths are defined as: $d_{SRk}$ (from S to
each RIS), $d_{SU}$ (direct path from S to U), and $d_{RUk}$ (from
each RIS to U). Given the global reach of satellite communications
and taking into account the Earth's spherical geometry, we apply spherical
trigonometry to determine the distances associated with signal propagation
\cite{Vallado2001}. The Cartesian coordinates of each RIS ($x_{rk},y_{rk},z_{rk}$)
and the user $(x_{u},y_{u},z_{u}$) are utilized to derive their corresponding
geographic coordinates (latitude and longitude). The coordinates for
the RIS are determined as:

\begin{gather}
x_{rk}=R\cos(\text{lat}_{r_{k}})\cos(\text{lon}_{r_{k}})\nonumber \\
y_{rk}=R\cos(\text{lat}_{r_{k}})\sin(\text{lon}_{r_{k}})\nonumber \\
z_{rk}=R\sin(\text{lat}_{r_{k}}),\label{eq: Xr}
\end{gather}

where the $k$ is an index that uniquely identifies each RIS and is
defined as $k\in(1,2...K)$, where $K$ represents the total number
of RISs in the system. The terms lat and lon denote the latitude and
longitude of the respective locations, and $R$ signifies the Earth's
radius. By employing the Cartesian coordinates of each RIS and the
user, we can determine the Euclidean distance $d_{RUk}$ between each
RIS and the user receiver 

\begin{equation}
d_{RUk}=\sqrt{(x_{rk}-x_{u})^{2}+(y_{rk}-y_{u})^{2}+(z_{rk}-z_{u})^{2}.}
\end{equation}

To deduce the total power received from both the direct LOS link and
any RISs, we commence by evaluating the power of the LOS signal arriving
at the ground receiver, denoted as U. It is assumed that all the channel
information from each RIS and the LOS link is available at the user
node. The power received by U is influenced by factors such as the
propagation distance $d_{\mathrm{SU}}$, the transmission frequency,
and atmospheric attenuation. Consequently, the expression which encapsulates
the complex amplitude of the signal from the satellite to the user,
incorporating both amplitude and phase information, is articulated
by

\begin{equation}
S_{\mathrm{SU}}=\sqrt{\frac{P_{\mathrm{t}}G_{\mathrm{t}}G_{\mathrm{U}}}{L_{SU}}}e^{-j\phi_{\mathrm{SU}}}.
\end{equation}

In the given expression, $P_{\mathrm{t}}$ denotes the transmitted
signal power, $L_{SU}$ is the path loss (PL) between S and U, while
$G_{\mathrm{t}}$ and $G_{\mathrm{U}}$ represent the gains of the
transmitter and receiver antennas, respectively. The phase of the
direct received signal is represented by $\phi_{\mathrm{SU}}$, and
is defined as

\begin{equation}
\phi_{\mathrm{SU}}=\frac{2\pi}{\lambda}d_{\mathrm{SU}},\label{eq: Ssu}
\end{equation}

The path loss $L_{SU}$ between S and U primarily arises from free
space propagation, and can be expressed in decibels using the following
expression:
\begin{equation}
\mathit{L_{\mathrm{SU}}\left[\mathrm{dB}\right]=\mathrm{20log(\mathit{f_{c}})+20log_{10}\mathit{(d_{\mathrm{SU}}})}-\mathrm{147.55}+\eta_{\mathrm{SU}},}
\end{equation}

where $\mathit{f_{c}}$ is the frequency in $Hz$, $\lambda$ is the
wavelength, and $\eta_{\mathrm{SU}}$ is a random variable captures
the additional PL arising from ground clutter, including structures
like buildings, vegetation, and terrain \cite{ITUR2021}. It's noteworthy
that when considering the RIS, which typically maintains a clear path
to the satellite, the shadowing effect (or additional PL) due to such
obstacles is negligible. The direct signal path from $\mathrm{S}$
to the user, $\mathrm{U}$, is characterized by both amplitude and
phase, represented as:

\begin{equation}
S_{\mathrm{SU}}=A_{\mathrm{SU}}e^{-\phi_{\mathrm{SU}}},
\end{equation}

where $A_{\mathrm{SU}}$ is given by: $A_{\mathrm{SU}}=\sqrt{P_{t}G_{\mathrm{u}}G_{\mathrm{t}}/L_{\mathrm{SU}}}$
and is accompanied by its phase shift $\phi_{\mathrm{SU}}.$ 

The trajectory of a signal incident on the $k^{th}$ RIS is is primarily
determined by the distance $d_{\mathrm{SR\mathit{k}}}.$ For each
$k^{th}$ RIS in the system (where $k$ ranges from 1 to $K$), the
incident signal can be comprehensively described by: \textbf{
\begin{equation}
S_{SRi}=\sqrt{P_{t}G_{\mathrm{t}}\mathrm{\mathrm{G(\varphi_{\mathit{\overset{i}{kn}}})}}/L_{\mathrm{SR\mathit{k}}}}e^{-j\phi_{\mathrm{SR\mathit{k}}}}.
\end{equation}
}

Here, $G(\varphi_{\mathit{\overset{i}{kn}}})$ denotes the antenna
gain at the $\mathit{n}$$^{th}$ element of the $k^{th}$ RIS, corresponding
to the angles of incidence $\varphi_{\mathit{\overset{i}{kn}}}$.
The phase $\phi_{\mathrm{SR\mathit{k}}}$ characterizes the phase
of the signal upon reaching the $k^{th}$ RIS and is defined as: $\phi_{SR}=\frac{2\pi}{\lambda}d_{\mathrm{SR}k}.$
Furthermore, $L_{\mathrm{SR\mathit{k}}}$ represents the path loss
between the S and the specific $k^{th}$ RIS. When expressed in decibels,
this path loss is given by:\textbf{
\begin{equation}
L_{\mathrm{SRk}}\left[\mathrm{dB}\right]=20\log_{10}(f_{\mathrm{c}})+20\log_{10}(d_{\mathrm{SRk}})-147.55+\eta_{\mathrm{SR\mathit{k}}}.
\end{equation}
}

In this expression, $\eta_{\mathrm{SR\mathit{k}}}$ is defined to
have the same characteristics as $\eta_{\mathrm{SU}.}$

The interaction of a signal with a given RIS is determined by several
factors. These encompass the angle of incidence, signal polarization,
the distance from the S to a specific RIS, denoted as $d_{\mathrm{SR\mathit{k}}}$,
and the distance from that RIS to the user receiver, indicated by
$d_{\mathrm{RU\mathit{k}}}$.

The complex amplitude of the signal reflected from any RIS, notated
as $S_{SR_{o}}$, is determined by:

\begin{equation}
S_{SR_{o}}=\sum_{k=1}^{K}\left(S_{SR_{i}}\sum_{n=1}^{N}\sqrt{G(\varphi_{i_{kn}})G(\varphi_{o_{kn}})}\Gamma_{kn}e^{-j\varphi_{kn}}\right).
\end{equation}

In this formulation, $\varphi_{kn}$ signifies the phase induced by
each element of the RISs, $G(\varphi_{\mathit{\overset{o}{kn}}})$
represents the reflected antenna gain of the $\mathit{n}^{th}$ element
of the RISs, correlated to the reflected angle $\varphi_{\mathit{\overset{o}{kn}}},$and
$\Gamma_{kn}$ illustrates the reflection coefficient for an RIS,
indicating the portion of the incident signal's power that's reflected
by the $\mathit{n}^{th}$ element within each RIS, as discussed in
\cite{Khalil2022}. The phase-shifting capability of each RIS panel,
denoted by $\varphi_{kn}$, plays a pivotal role in the system. This
capability to modulate the signal phase can induce either constructive
or destructive interference at the receiving terminal. Under the assumption
that there are no multi-path signals originating from any RIS to the
receiver, the complex amplitude of the signal received by U is given
by:

\begin{gather}
S_{\mathrm{RU\mathit{k}}}=\sum_{\mathrm{\mathit{k}}=1}^{K}\left(\sum_{n=1}^{N}\sqrt{\frac{P_{t}G(\varphi_{\mathit{\overset{i}{kn}}})G(\varphi_{\mathit{\overset{o}{kn}}})G_{\mathrm{t}}}{L_{\mathrm{SR_{\mathit{k}}}}}}\right.\nonumber \\
\left.\Gamma_{kn}\sqrt{\frac{G_{\mathrm{U}k}}{L_{\mathrm{RU\mathit{k}}}}}e^{-j\left(\phi_{\mathrm{RU\mathit{k}}}+\phi_{\mathrm{SR_{\mathit{k}}}}+\varphi_{kn}\right)}\right),\label{eq:a}
\end{gather}

where $\phi_{\mathrm{RU\mathit{k}}}$ represents the phase of the
indirectly received signal and is defined by the equation $\phi_{\mathrm{RU\mathit{k}}}=\frac{2\pi}{\lambda}d_{\mathrm{RU\mathit{k}}},$
and $L_{\mathrm{RU\mathit{k}}}$ , denotes the path loss between a
specific RIS and the user U and can be expressed as

\begin{equation}
L_{\mathrm{RU\mathit{k}}}\left[\mathrm{dB}\right]=20\log_{10}(f_{\mathrm{c}})+20\log_{10}(d_{\mathrm{RUk}})-147.55+x_{k}.
\end{equation}

In this equation, $x_{k}$ is a random variable representing the shadow
fading effect between any RIS and the user. This shadow fading is
characterized by a log-normal distribution: $X_{k}\sim\mathrm{\log}(\mu_{\mathrm{RU\mathit{k}}},\sigma_{\mathrm{RU\mathit{k}}})$,
where , $\mu_{\mathrm{RU\mathit{k}}}$ and $\sigma_{\mathrm{RU\mathit{k}}}$
are the mean and standard deviation of the shadow fading effect for
any $RIS$, respectively. Drawing upon the amplitude expression from
\eqref{eq:a}, we derive 

\begin{equation}
A_{kn}=\left|\sqrt{\frac{G(\varphi_{\mathit{\overset{i}{kn}}})G(\varphi_{\mathit{\overset{o}{kn}}})G_{\mathrm{t}}G_{\mathrm{U}k}}{L_{\mathrm{SR_{k}}}L_{\mathrm{RU_{k}}}}}\Gamma_{kn}\right|,
\end{equation}

with the total phase being: $\phi_{kn}=\phi_{\mathrm{RUk}}+\phi_{\mathrm{SR_{k}}}+\varphi_{kn}$. 

Now, simplifying the expression for $S_{\mathrm{RU_{\mathit{k}}}}$
we get: 

\begin{equation}
S_{\mathrm{RU}_{k}}=\sqrt{P_{t}}\sum_{k=1}^{K}\sum_{n=1}^{N}A_{kn}e^{-j\left(\phi_{kn}\right)},\label{eq:Su}
\end{equation}

Let us define the overall amplitude of the signal at the user from
all the RIS elements as: $A_{RU_{k}}=\sqrt{P_{t}}\sum_{k=1}^{K}\sum_{n=1}^{N}A_{kn},$and
the phase of the total signal at the user from all the RIS elements
as: 
\begin{equation}
\Theta_{total}=\arg(S_{\mathrm{RU\mathit{k}}}),
\end{equation}

where the function $\arg(.)$ extracts the phase angle of a complex
number. This $\Theta_{total}$ essentially captures the overall phase
shift experienced by the signal as it interacts with all the RIS elements
and reaches the user. However, to delve deeper and understand the
phase contributions from each individual RIS element, we can represent
these individual phases in a vector format. For each RIS element,
indexed by $k$ and $n$ , the phase is given by $\phi_{kn}$ . Collectively,
these phases can be represented as a vector: 

\begin{equation}
\Theta=\left[\phi_{11},\phi_{kn},...,\phi_{KN}\right],\label{eq:14}
\end{equation}

Each element in the $\Theta$ vector represents the phase shift introduced
by a specific RIS element. By examining this vector, we can glean
insights into how each RIS element contributes to the overall phase
of the received signal.

The total signal received at the user is the summation of both the
direct path from the satellite and the reflected paths from all the
RISs, represented as $\left|S_{\mathrm{SU}}+S_{\mathrm{RU\mathit{k}}}\right|.$
The received power, $P_{R}$, is then determined by squaring the magnitude
of this combined signal:

\begin{equation}
P_{R}=\left|S_{\mathrm{SU}}+S_{\mathrm{RU\mathit{k}}}\right|^{2}.\label{eq:12}
\end{equation}

For any complex number $z$, the squared magnitude is given by,$\left|z\right|^{2}=z\:\overset{*}{z}$,
where $\overset{*}{z}$ denotes the complex conjugate of $z.$ Utilizing
this property, the received power can be expanded as:

\begin{equation}
P_{R}=\Bigl(S_{\mathrm{SU}}+S_{\mathrm{RU\mathit{k}}}\Bigr)\Bigl(\overset{*}{S_{\mathrm{SU}}}+\overset{*}{S_{\mathrm{RU\mathit{k}}}}\Bigr).\label{eq:term_PR}
\end{equation}

Breaking down the terms in \eqref{eq:term_PR} as 

\begin{gather}
P_{R}=\underset{\underset{\mathrm{RIS-reflected\:signal}}{\mathrm{power\,}\mathrm{due\:}\mathrm{to}}}{\overset{*}{\underbrace{S_{\mathrm{RU\mathit{k}}}S_{\mathrm{RU\mathit{k}}}}}}+\underset{\underset{\mathrm{direct\:signal}}{\mathrm{power\:}\mathrm{due\,}\mathrm{to}}}{\overset{}{\underbrace{S_{\mathrm{SU}}\overset{*}{S_{\mathrm{SU}}}}}}+\underset{\underset{\mathrm{interferencel\:}\mathrm{between}\mathrm{\:signals}}{\mathrm{cross-terms}\mathrm{\:represnting}}}{\underbrace{S_{\mathrm{RU\mathit{k}}}\overset{*}{S_{\mathrm{SU}}}+S_{\mathrm{SU}}\overset{*}{S_{\mathrm{RU\mathit{k}}}}},}\label{eq:R_P}
\end{gather}

where $\overset{*}{S_{\mathrm{RU\mathit{k}}}}S_{\mathrm{SU}}$ and
$\overset{*}{S_{\mathrm{SU}}}S_{\mathrm{RU\mathit{k}}}$ represent
the cross-terms, indicating the interaction between the direct signal
and the RIS-reflected signal. Using properties of complex multiplication,
this interaction is defined as: $S_{\mathrm{SU}}\overset{*}{S_{\mathrm{RU\mathit{k}}}=}A_{RU_{k}}A_{\mathrm{SU}}e^{-j\left(\phi_{\mathrm{SU}}-\Theta\right)}.$
These cross-terms encapsulate the interference effects between the
direct and RIS-reflected signals, accounting for both amplitude and
phase differences. In particular, the real part of this product captures
the nature of this interference, signifying whether the signals constructively
or destructively interfere. Hence, the received power can be succinctly
represented as:

\begin{equation}
P_{R}=P_{t}\left[A_{RU_{k}}^{2}+A_{\mathrm{SU}}^{2}+2A_{RU_{k}}A_{\mathrm{SU}}\cos\left(\phi_{\mathrm{SU}}-\Theta\right)\right].\label{eq: P(R)}
\end{equation}

Thus, the received power $P_{R}$ is a function of various factors,
including phase shifts, propagation distances, and external environmental
influences represented by path loss.

\section{Energy Efficiency \label{sec:EE } }

The incorporation of RISs into contemporary wireless communication
architectures has been identified as a promising avenue for bolstering
overall system performance. In scenarios where the system comprises
$K$ distinct RISs, each equipped with $N$ reflecting elements, the
metric of energy efficiency becomes crucial for a comprehensive assessment
of the system's effectiveness \cite{Yang2022a}. As delineated by
\cite[p. 2484]{Whitaker2005}, energy efficiency is represented as
the ratio of the total power received by the system to its overall
power consumption. In formal terms, energy efficiency can be expressed
as:

\begin{equation}
\eta=\frac{P_{R}}{P_{t}+K\,P_{\text{crt}}+\ensuremath{\sum\limits _{k=1}^{K}\sum\limits _{n=1}^{N}s_{kn}\bigl(P_{\text{el},kn}+P_{\text{con,}kn}\bigr)}}.\label{eq: EE}
\end{equation}

In \eqref{eq: EE} formulation, the term $\ensuremath{P_{\text{crt}}}$
represents the power consumed by each RIS system's circuitry. Assuming
all the RISs have approximately the same $P_{\text{crt}}$, then $K\,P_{\text{crt}}$.
The variable $s_{kn}$ denotes the state of the $n^{th}$ element
in the $k^{th}$ RIS. Specifically, $\ensuremath{s_{kn}=1}$ signifies
that the $n^{th}$ reflecting element in the $k^{th}$ RIS is active,
whereas $s_{kn}=0$ indicates it is inactive. The terms $P_{\text{el},kn}$
and $P_{\text{con,}kn}$ indicate the power consumed for phase shifting
and control operations, respectively, for the $n^{th}$ element of
the $k^{th}$ RIS. The term $P_{t}+K\,P_{\text{crt}}+\ensuremath{\sum\limits _{k=1}^{K}\sum\limits _{n=1}^{N}s_{kn}\bigl(P_{\text{el},kn}+P_{\text{con,}kn}\bigr)}$
encapsulates the notion that power consumption for phase shifting
and control may vary across elements and RISs. In our study, equation
$\eqref{eq: EE}$ serves as the reference or baseline model, representing
the standard approach in the field. Within this model, the term \eqref{eq: P(R)}
is defined as per \eqref{eq: P(R)}. Notably, this baseline does not
account for advanced techniques such as phase optimization, or the
implementation of selective diversity.

\begin{figure}
\centering{}\includegraphics[viewport=3bp 0bp 540bp 340bp,width=3.5in]{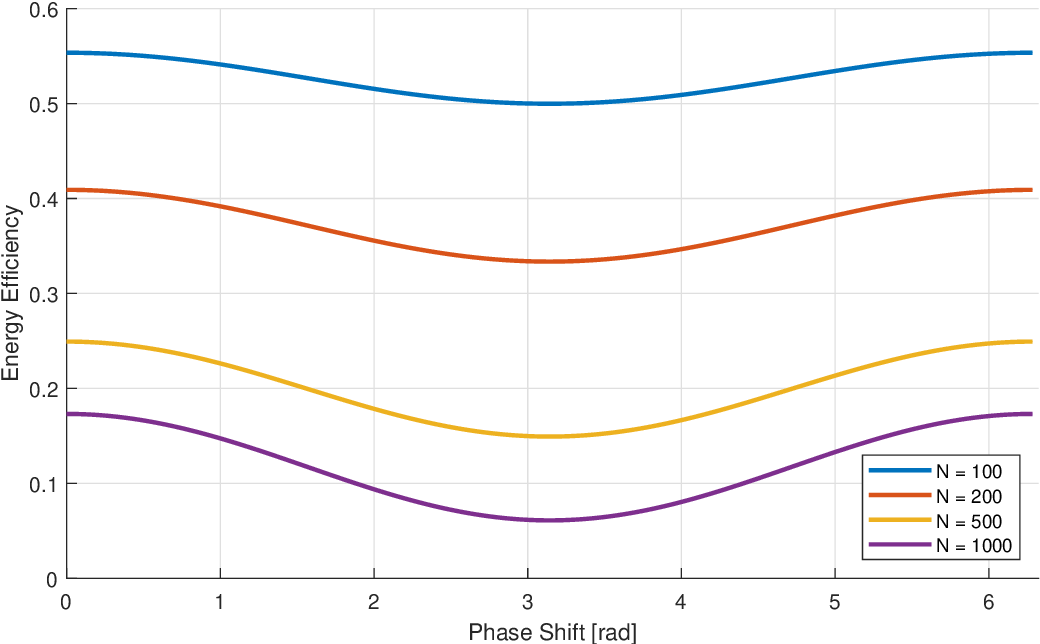}\caption{Baseline energy efficiency for multiple RIS as a function of phase
shift range: associating each curve with a specific $N$ Number \label{fig: 1}}
\end{figure}

Fig. \ref{fig: 1} depicts the baseline energy efficiency of a system
that incorporates four RISs. For the sake of simplification, we assume
that all RIS elements have a uniform phase shift. The graph indicates
a decline in energy efficiency as the number of reflecting elements,
$N$, increases. This decline can be attributed to the increased power
consumption by the RISs, especially during phase shifting and the
control of additional elements. Consequently, this amplifies the denominator
in the \textbf{$\eta$ }equation, leading to a reduction in overall
energy efficiency. Such observations emphasize the inherent trade-off
between system performance and energy consumption. It underscores
the importance of judiciously selecting the number of reflecting elements
in RIS-integrated communication systems, even before the implementation
of any optimization techniques.

Subsequently, our primary aim is to optimize energy efficiency, which
intrinsically reduces power consumption while concurrently enhancing
received power. To achieve this, we deconstruct \eqref{eq: EE} into
two sub-optimal problems: maximization of received power $(i.e.,P_{R})$
and minimization of power consumption, $\ensuremath{P_{t}+K\,P_{\text{crt}}+\ensuremath{\sum\limits _{k=1}^{K}\sum\limits _{n=1}^{N}s_{kn}\bigl(P_{\text{el},kn}+P_{\text{con,}kn}\bigr)}}$.
In this study, $\eta$ maximization is introduced into two scenarios.
The first scenario considered ideal conditions, assuming no path loss,
while the second one incorporates the path loss.

\section{Maximize {\normalsize{}$\eta$} under ideal conditions \label{ideal } }

The primary objective of the proposed method is to optimize the phase
shift, \textbf{$\varphi_{kn}$,} to enhance $\eta$ over a specified
distance range$,d_{\mathrm{RU}}$. Given that \textbf{$\varphi_{kn}$}
is an element of the phase vector $\Theta$, and considering that
other components (i.e., $\phi_{\mathrm{RUk}}$ and $\phi_{\mathrm{SR_{k}}}$)
are known, the main focus shifts to the optimization of \textbf{$\Theta\LyXZeroWidthSpace$}.
By achieving this, we can effectively cover a broader range of users
or areas. As previously mentioned, maximizing $\eta$ entails maximizing
$P_{R}$ and minimizing the denominator of $\eta$ . 

The received power, $P_{R}$, depends on several factors, including
the transmitted power, the gains of the transmitting and receiving
antennas, the path loss, and the phase shifts introduced by the RIS
elements. In the absence of path loss, the received power is predominantly
influenced by the direct LOS path and the paths associated with the
RISs. The interference effect between the direct and RIS-reflected
signals is encapsulated by the term $\cos\left(\phi_{\mathrm{SU}}-\Theta\right)$.
The goal is to adjust the phase shifts $\Theta$ to ensure this interference
is constructive, thereby augmenting the received power. To identify
the optimal phase shifts that maximize $P_{R},$ we employ differential
calculus. Differentiating $P_{R}$ with respect to $\Theta$ provides:

\begin{equation}
\frac{\partial P_{R}}{\partial\Theta}=-2P_{t}A_{RU_{k}}A_{SU}\sin(\phi_{SU}-\Theta)\label{eq: 18}
\end{equation}

To maximize $P_{R}$, the derivative should be set to zero, yielding
the condition: $\phi_{SU}-\Theta=n\pi$, where, $n$ is an integer.
This equation delineates the optimal phase shifts that ensure the
received power is at its peak. Essentially, the phase shifts must
be adjusted so that the direct and RIS-reflected signals are in phase.
This alignment results in constructive interference, leading to an
enhanced received power.

\begin{figure}
\centering{}\includegraphics[width=3.5in]{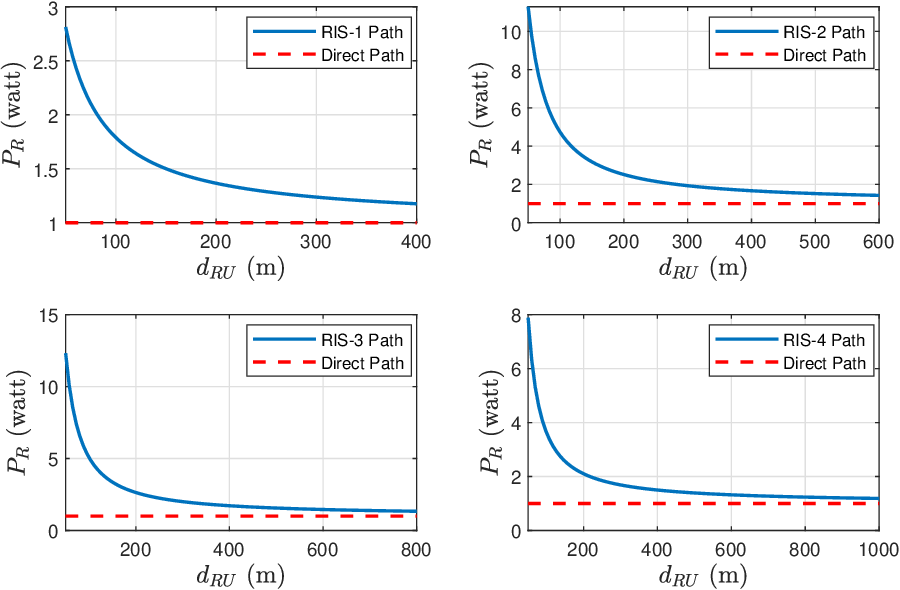}\caption{Comparative analysis of received power via direct path and multiple
RISs across varied distances\label{fig: 2 }}
\end{figure}

To validate \eqref{eq: 18}, we assess the power received from each
individual RIS, treating each case as if $K=1$. Fig. \ref{fig: 2 }
presents a detailed comparison of the received power when utilizing
four RISs versus a direct path over a range of distances. Each subplot
is associated with a distinct RIS, differentiated by its proximity
to the user. The data suggests that the received power is influenced
by several factors, such as the distance between the RIS and the user,
the number of elements in each RIS, and the phase shifts introduced
by the RIS. The red dashed line in each subplot represents the power
received via the direct path, serving as a benchmark for comparison.
On the other hand, the continuous lines illustrate the power received
with the aid of each RIS. A side-by-side comparison of the continuous
and dashed lines clearly demonstrates the potential of an RIS to either
enhance or reduce the received power relative to the direct transmission
path.

Given a system with multiple paths (or channels) through which a signal
can reach a receiver, represented as $\left(P_{R1},P_{R1},..,P_{RK}\right),$
selective diversity can be employed to choose the path with the highest
signal power. This can be mathematically expressed as:

\begin{equation}
P_{Rk}=max\left\{ P_{R1},P_{R1},..,P_{RK}\right\} \label{eq:Max_idel}
\end{equation}

where $P_{Rk}$ represents the selected power among $K$ RISs. 

The secondary objective focuses on minimizing the cumulative power
consumption, represented as: $P_{t}+K\,P_{\text{crt}}+\ensuremath{\sum\limits _{k=1}^{K}\sum\limits _{n=1}^{N}s_{kn}\bigl(P_{\text{el},kn}+P_{\text{con,}kn}\bigr)}$.
To augment power efficiency across the circuitry and individual RIS
elements, several pivotal tactics warrant consideration. One such
strategy entails the selective deactivation of certain RIS elements
deemed non-essential, facilitating power conservation. Furthermore,
meticulous scrutiny and optimization of control operations can yield
energy-efficient management, curtailing power expenditure without
sacrificing system efficacy. On a more granular level, hardware-centric
enhancements offer avenues for superior energy efficiency. This can
be achieved by diminishing the intrinsic power consumption of each
RIS element, either through the adoption of energy-conserving materials
or advancements in thermal management. Consequently, the optimization
objective can be articulated as:

\begin{gather}
\text{minimize}\quad\left(P_{t}+P_{\text{crt}}+\ensuremath{\sum\limits _{k=1}^{K}\sum\limits _{n=1}^{N}s_{kn}\bigl(P_{\text{el},kn}+P_{\text{con,}kn}\bigr)}\right)\nonumber \\
\text{s.t.}\quad s_{kn}\in\{0,1\}\quad\forall k\in\{1,2,\ldots,K\},\forall n\in\{1,2,\ldots,N\}.\label{eq: min}
\end{gather}

The primary objective now is purely to reduce the power consumed by
the RISs. Since $s_{kn}$ is binary, the optimization will find the
optimal configuration of active and inactive reflecting elements to
achieve this objective. The problem defined in \eqref{eq: min} is
characterized as a binary linear programming issue. While there exist
efficient solvers tailored for such challenges, the computational
burden significantly rises with an increase in the number of RISs
$K$ and elements number $N.$ As a result, resorting to heuristic
or metaheuristic strategies, such as greedy algorithms and genetic
algorithms, could offer a more feasible approach. However, Particle
Swarm Optimization (PSO) stands out among these methods due to its
unique ability to concurrently explore and exploit the solution space.
Its collaborative nature enables particles to learn from both individual
and global best solutions, ensuring rapid convergence. Moreover, when
adapted for binary contexts using Binary PSO (BPSO) techniques, PSO
demonstrates enhanced scalability and efficiency with minimal parameter
tuning, marking it as an optimal choice for such optimization challenges
\cite{Cervante2012}.

Given $K$ RISs, where each RIS has $N$ reflecting elements, we can
represent the state $s_{kn}$ (either active or inactive) of each
element using a binary matrix. The dimensionality of our swarm is
determined by the total number of variables in our problem, which
is \textbf{$K\times N$}. In the PSO approach, every particle embodies
a potential solution. The position of a particle is represented by
a binary matrix, indicating whether each reflecting element is active
or inactive. To illustrate, the binary state of reflecting elements
across all RISs can be captured in the matrix $S$ as follows:

\begin{equation}
S=\begin{bmatrix}s_{11} & s_{12} & \dots & s_{1N}\\
s_{21} & s_{22} & \dots & s_{2N}\\
\vdots & \vdots & \ddots & \vdots\\
s_{K1} & s_{K2} & \dots & s_{KN}
\end{bmatrix},\label{eq: Matrix}
\end{equation}

where $s_{kn}\in\{0,1\}$ denotes the state of the $n^{th}$ element
in the $k^{th}$ RIS. 

The aim of the PSO algorithm is to identify the matrix $S$ that results
in the least total power consumption. This consumption is quantified
by the fitness function $F(S)$ as:

\begin{equation}
F(S)=P_{t}+K\,P_{\text{crt}}+\sum\limits _{k=1}^{K}\sum\limits _{n=1}^{N}s_{kn}\bigl(P_{\text{el},k}+P_{\text{con,}k}\bigr).
\end{equation}

In this context, each particle possesses both a position and velocity.
The position defines the current solution, representing the activation
configuration of the RIS elements. On the other hand, the velocity
dictates the particle's positional shift in the succeeding iteration.
The velocity update equation for a given reflecting element is defined
as:

\begin{equation}
v_{kn}^{(t+1)}=w\cdot v_{kn}^{(t)}+c_{1}\cdot r_{1}\cdot(p_{kn}^{best}-s_{kn}^{(t)})+c_{2}\cdot r_{2}\cdot(g_{kn}^{best}-s_{kn}^{(t)})
\end{equation}

where, $v_{kn}^{(t+1)}$ is the velocity of the element $kn$ at time
$t+1$, $w$ is the inertia weight which controls the impact of the
previous velocity, $s_{kn}^{(t)}$ represents the position (or state)
of the element $kn$ at time $t$, $c_{1}$ and $c_{2}$ are cognitive
and social constants, respectively, dictating how much the particle
considers its own best position and the best position of its neighbors,
$r_{1}$ and $r_{2}$ are random numbers between 0 and 1, introducing
a stochastic aspect to the optimization, $p_{kn}^{best}$ is the best-known
position for that element, g$_{kn}^{best}$ is the best-known position
among the particle's neighbours. 

After updating the velocity, it's necessary to determine if the reflecting
element is active or inactive. To convert the continuous velocity
value into a binary choice, we utilize the sigmoid transfer function,
given by 

\begin{equation}
\sigma(v)=\frac{1}{1+e^{-v}}
\end{equation}

The sigmoid function translates the velocity into a value between
0 and 1, aiding in deciding the state of the reflecting element. If
the result of the sigmoid function, $\sigma(v_{kn}^{(t+1)})$, exceeds
0, the element is designated as active (1); otherwise, it remains
inactive (0). Therefore, the updated position can be defined as:

\begin{equation}
s_{kn}^{(t+1)}=\begin{cases}
1 & \text{if }\sigma(v_{kn}^{(t+1)})>0.5.\\
0 & \text{otherwise}
\end{cases}
\end{equation}

The threshold \textquotedbl$>0.5$\textquotedbl{} is used for balanced
binary decision-making. Through the sigmoid function, values greater
than 0.5 yield a decision of 1, while those less or equal give 0.
This offers a symmetrical and unbiased division of possible outcomes.

The algorithm continues either for a predetermined number of iterations
or until convergence is achieved, with the fitness function reaching
its minimal value. The resulting configuration matrix identifies the
best activation states for the RIS reflecting elements, aiming to
curtail power consumption. To incorporate the activation status of
each RIS element into the power consumption model, the formula must
be adjusted as 

\begin{equation}
P_{t}+K\,P_{\text{crt}}+\sum_{k=1}^{K}\sum_{n=1}^{N}s_{kn}^{(t+1)}\bigl(P_{\text{el},k}+P_{\text{con,}k}\bigr)\label{eq:33}
\end{equation}

The expression \eqref{eq:33}, captures the overall power consumption
of the system at a specific iteration $t+1$ in the optimization process.
Here, the state $s_{kn}^{(t+1)}$ signifies whether an individual
RIS element is active or not at the $t+1$ iteration. When an RIS
element is active, represented by $s_{kn}^{(t+1)}=1$, it utilizes
power for both phase adjustment and control tasks. Conversely, if
the RIS element is inactive, indicated by $s_{kn}^{(t+1)}=0$, it
refrains from consuming power for these functions. By the conclusion
of the optimization procedure, employing the BPSO algorithm, each
$s_{kn}^{(t+1)}$ will finalize to a value of either 0 or 1. This
indicates the optimal operational state of every RIS element to minimize
the system's total power consumption. The $t+1$ notation is indicative
of the progression step within the optimization process. As the algorithm
advances, it systematically updates the operational states of the
RIS elements, based on the inherent velocity and position update rules
of the BPSO methodology. The ultimate objective is to discern the
configuration of RIS elements that curtails the total power consumption
while preserving optimal system performance. 

\begin{figure}
\centering{}\includegraphics[width=3.5in]{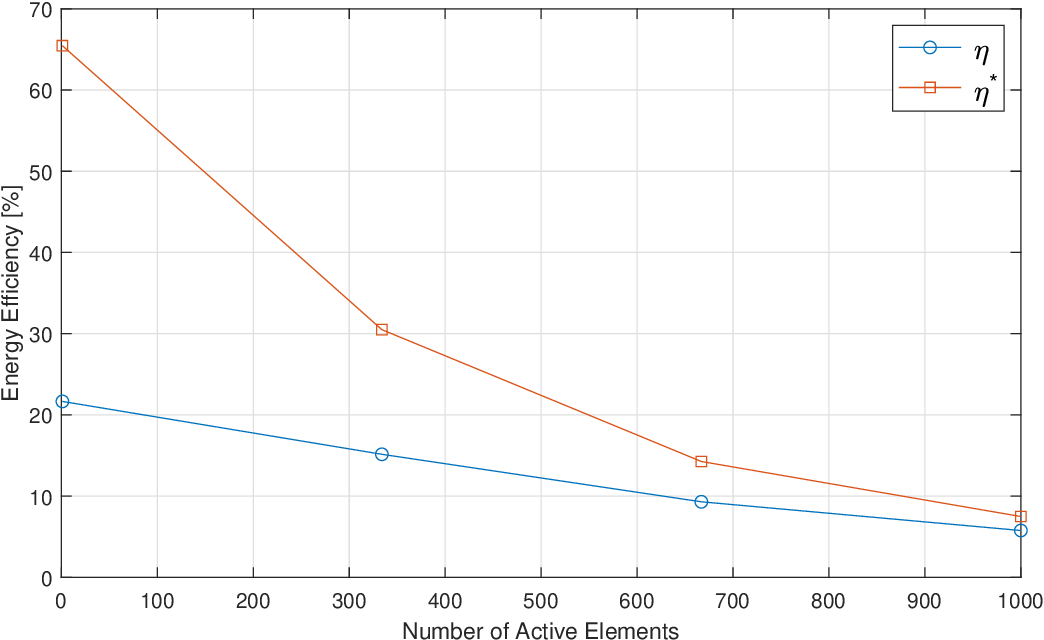}\caption{Energy efficiency comparison between baseline and optimized methods
across active reflecting elements \label{fig: 4 }}
\end{figure}

Incorporating \eqref{eq:Max_idel} and \eqref{eq:33}, we can deduce
the optimal energy efficiency for the proposed system. Comparing this
efficiency with the baseline $\eta$ outlined in \eqref{eq: EE},
Fig. \ref{fig: 4 } illustrates the energy efficiency in relation
to the number of active elements for both the baseline and the optimized
scenarios. The comparison underscores that the optimal $\eta$ ($\overset{*}{\eta}$)
delivers a markedly better energy efficiency than its baseline counterpart,
underlining the potency of our advocated technique.

\begin{figure}
\centering{}\includegraphics[width=3.5in]{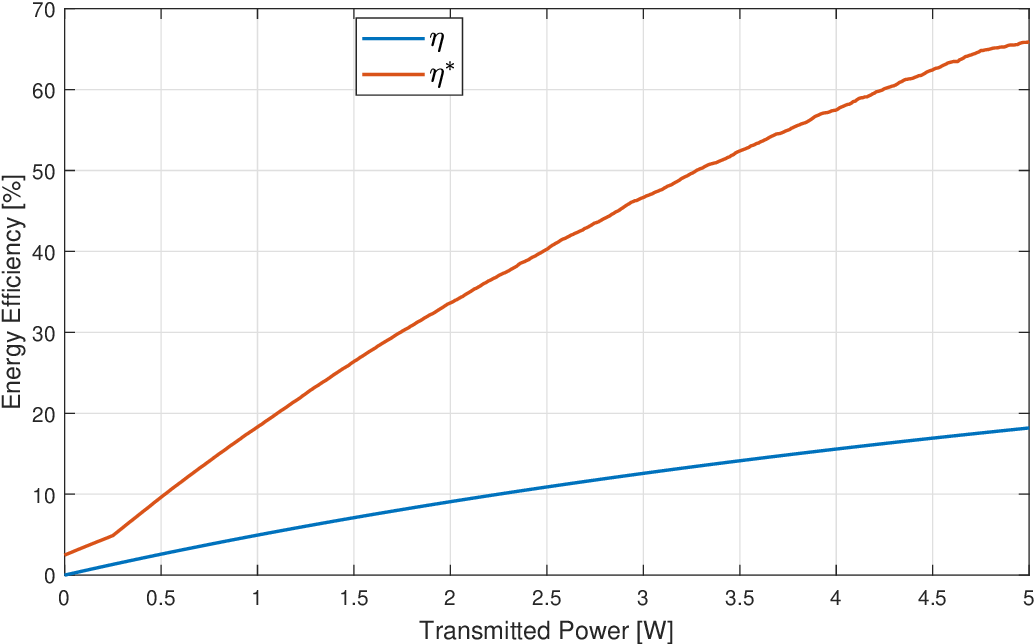}\caption{Comparison of Energy efficiency between baseline and optimized methods
across transmitted power levels \label{fig: 5 }}
\end{figure}

In practical wireless communication systems, transmitters dynamically
adjust their power output in response to various factors, including
environmental conditions and channel states. Such modifications can
significantly influence the system's energy efficiency, underscoring
the importance of comprehending its implications. Fig. \ref{fig: 5 }
illustrates the relationship between transmitted power (denoted as
$P_{t}$) and energy efficiency within the proposed systems. For analytical
purposes, two scenarios are presented: a baseline model devoid of
optimization techniques and an enhanced method, $\overset{*}{\eta}$,
which implements the proposed optimization strategy to augment energy
efficiency. Energy efficiency is measured at different transmitted
power levels, showing how $P_{t}$ affects energy efficiency in both
scenarios. The discernible disparities between the curves underscore
the benefits of integrating optimization techniques, showcasing superior
energy efficiency across different transmitted power levels in the
optimized method compared to the baseline.

\section{Maximize {\normalsize{}$\eta$} considering non-ideal condition \label{sec: EE_non idel} }

In the complex realm of wireless communication systems, path loss
significantly influences performance, particularly in suboptimal environmental
conditions. Path loss inherently encompasses deterministic factors,
such as distance and environmental attributes, as well as stochastic
components, primarily arising from shadowing effects denoted by $x_{k}.$
Considering this variability, instead of aiming for the highest possible
energy efficiency, the goal becomes maximizing its average value as
follows:

\begin{gather}
\mathbb{E}[\eta(\Theta,L_{\mathrm{RU\mathit{k}}})]=\nonumber \\
\mathbb{E}\left[\frac{P_{R}(\Theta,L_{\mathrm{RU\mathit{k}}})}{P_{t}+K\,P_{\text{crt}}+\sum\limits _{k=1}^{K}\sum\limits _{n=1}^{N}s_{kn}\bigl(P_{\text{el},k}+P_{\text{con,}k}\bigr)}\right].\label{eq:=00005BExpect=00005D}
\end{gather}

In the equation provided, $\mathbb{E}[\eta(\Theta,L_{\mathrm{RU\mathit{k}}})]$
signifies the expected value, which is a function of $\Theta$ and
a random variable. The path-loss, $L_{\mathrm{RU\mathit{k}}}$ in
a linear scale . To achieve \eqref{eq:=00005BExpect=00005D}, which
includes both direct LOS signals and those reflected by RISs, we utilize
the Adam optimization algorithm. This method is adept at managing
intricate optimization challenges, rendering it appropriate for our
system. Our system integrates various elements, from direct signals
and RIS-reflected signals to phase alterations and shadowing phenomena.
A notable characteristic of the Adam method is its adaptive learning
rates. These rates adjust based on historical gradient data \cite{Kingma2017}.
This flexibility provides stability against unpredictable variations
introduced by shadowing $x_{k}$, ensuring consistent convergence
of the algorithm. Moreover, the momentum feature of Adam aids in navigating
challenges in the optimization landscape, such as local extremes or
unstable points, particularly those arising from interactions between
direct and RIS-reflected signals. The RIS\textquoteright s capability
to alter phase is central to our system and benefits from the adaptability
of Adam, ensuring optimal phase values within predefined limits. In
light of these advantages, the Adam algorithm maintains computational
efficiency even as our system expands to include multiple RISs, each
with numerous reflecting elements. This underscores its role as a
reliable tool for enhancing the energy efficiency of our system.

To optimize the system using the Adam algorithm, the process commences
with the random initialization of phase shifts, $\varphi_{kn}$, for
each RIS element, ensuring they are constrained within the range {[}0,$\:$2\textgreek{p}{]}.
This represents our starting point in the optimization landscape.
Following this, two moment estimators,$M_{t}$ and $V_{t}$ are initialized
to zero. These estimators are instrumental in adaptively adjusting
the learning rates for each specific parameter, $\varphi_{kn}$. During
the optimization, we track the progress using a time step, denoted
as''$t$'', which starts at zero and increments with each iteration.
To prevent division by zero in our calculations, we introduce a small
constant value, $\epsilon=1\times10^{-8}$, ensuring stability and
accuracy throughout the computational process. The core of the optimization
relies on the estimation of the gradient, which is expressed as:

\begin{equation}
g_{t}\LyXZeroWidthSpace=\nabla\Theta\mathbb{E}[\eta(\Theta,L_{\mathrm{RU_{k}}})],
\end{equation}

where $\varphi_{kn}\in\Theta$.

The gradient of $\mathbb{E}[\eta(\Theta,L_{\mathrm{RU_{k}}})]$ with
respect to $\Theta$ indicates the necessary direction and magnitude
of adjustment to optimize the expected energy efficiency. Direct computation
of this gradient can be challenging due to the evaluation of expectations
over the random variable $x_{k}$ embedded in $L_{\mathrm{RU_{k}}}$.
Such evaluations might require the resolution of complex integrals
or summations, especially when the system model encompasses multiple
stochastic elements. To circumvent these intricacies and to encapsulate
the effects of the stochastic variable $x_{k}$, the Monte Carlo sampling
technique is employed. This method facilitates the approximation of
the gradient by generating numerous samples of $x_{k}$ from its established
distribution. The equation for the Monte Carlo approximation of the
gradient is given by:

\begin{equation}
\frac{\partial E\left[\eta(\Theta,L_{\mathrm{RU_{k}}})\right]}{\partial\Theta}\approx\frac{1}{M}\sum_{m=1}^{M}\frac{\partial\eta(\Theta,L_{\mathrm{RU_{k}}}^{(m)})}{\partial\Theta}.\label{eq:29}
\end{equation}

Here, $\mathit{M}$ is the number of Monte Carlo samples generated
from the distribution of $x_{k}$, and $x_{k}^{(m)}$ is the $m^{th}$
samples of $x_{k}$. The approximation method outlined in \eqref{eq:29}
provides a feasible strategy to estimate intricate expectations influenced
by stochastic variables. As the sample size increases, this method
tends to converge towards the true value, striking a balance between
computational efficiency and statistical precision. In essence, it
harmonizes the trade-off between precision and feasibility. Utilizing
this approximation, the update rule for the phase shift $\Theta$
can be formulated as 

\begin{equation}
\Theta^{(t+1)}=\Theta^{(t)}+\alpha\frac{1}{M}\sum_{m=1}^{M}\frac{\partial\eta(\Theta,L_{\mathrm{RU_{k}}}^{(m)})}{\partial\Theta},
\end{equation}

where $\alpha$ is the learning rate.

To improve the optimization procedure, the Adam algorithm employs
moment estimates for adaptive adjustment of the learning rates. The
initial moment estimate, often termed the momentum term, is given
by:

\begin{equation}
M_{t}=\beta_{1}M_{t-1}+(1-\beta_{1})\frac{1}{M}\sum_{m=1}^{M}\frac{\partial\eta(\Theta,L_{\mathrm{RU_{k}}}^{(m)})}{\partial\Theta}.
\end{equation}

The subsequent moment estimate, often equated to the Root Mean Square
Propagation (RMSP) term, serves as a dynamic adjuster for the learning
rate. This dynamic adjustment enhances the algorithm's stability in
the face of gradient variations. Formally, the RMSP term \textbf{$V_{t}$}
is expressed as:

\begin{equation}
V_{t}=\beta_{2}V_{t-1}+(1-\beta_{2})\left(\frac{1}{M}\sum_{m=1}^{M}\frac{\partial\eta(\Theta,L_{\mathrm{RU_{k}}}^{(m)})}{\partial\Theta}\right)^{2},
\end{equation}

where $\beta_{1}$ and $\beta_{2}$ denote the exponential decay rates
for the first and second moment estimates, respectively, and they
are typically close to one, $V_{t}$ serves as an adaptive normalization
factor, adjusting the learning rate for each parameter $\Theta.$

The values $M_{t}$ and $V_{t}$ correspond to the first moment (mean)
and the second moment (uncentered variance) of the gradients, respectively.
When initialized at zero, these estimates can be biased towards zero,
especially during the initial iterations. Such bias can hinder the
optimization process, especially at its commencement. To counteract
this bias, the Adam algorithm introduces bias-corrected versions of
$M_{t}$ and $V_{t}$. The corrected estimates are computed as:

\begin{equation}
\stackrel{\curlywedge}{M_{t}}=\frac{M_{t}}{1-\beta_{1}^{t}}\mathrm{,and}\stackrel{\curlywedge}{V_{t}}=\frac{V_{t}}{1-\beta_{2}^{t}}.
\end{equation}

Utilizing these bias-corrected values, the update formula for the
phase shift $\Theta$ is defined as: 

\begin{equation}
\alpha_{t}=\frac{\alpha}{\sqrt{\stackrel{\curlywedge}{V_{t}}}+\epsilon}.
\end{equation}

With the inclusion of the bias-corrected estimates, the update rule
for the phase shift is

\begin{equation}
\Theta^{(t+1)}=\Theta^{(t)}+\alpha\frac{\stackrel{\curlywedge}{M_{t}}}{\sqrt{\stackrel{\curlywedge}{V_{t}}}+\epsilon}.\label{eq:38}
\end{equation}

The updated phase shifts, after each iteration, are subjected to a
clipping procedure to ensure they remain within the permissible range.
Mathematically, this can be represented as:

\begin{equation}
\LyXZeroWidthSpace\overset{*}{\Theta}=\text{clip}(\Theta^{(t+1)},0,2\pi).
\end{equation}

The function $\text{clip}(\Theta^{(t+1)},0,2\pi)$ guarantees that
$\Theta^{(t+1)}$ remains within the defined interval $[0,2\pi],$maintaining
the physical validity of the model. Once all the optimal individual
phase shifts $\overset{*}{\Theta}$ are obtained, they are combined
into a unified variable, represented as $\overset{*}{\Theta}$. It
is formulated as:

\begin{equation}
\overset{*}{\Theta}=\left[\overset{*}{\phi_{11}},\overset{*}{\phi_{kn}},...,\overset{*}{\phi_{K}}\right].
\end{equation}

The $\overset{*}{\Theta}$ vector acts as a consolidated representation
of the optimal phase shifts throughout the system. Utilizing this
vector simplifies the process of comparing energy efficiencies among
different RISs. The selection of the most efficient RIS is based on
the subsequent criteria:

\begin{equation}
k^{*}=\arg\max_{k\in\{1,2,\ldots,K\}}\eta\bigl(\overset{*}{\Theta}\bigr).\label{eq:39}
\end{equation}

This methodology facilitates the optimization of phase shifts for
individual RISs and simultaneously identifies the RIS that delivers
the highest energy efficiency. The step-by-step process of applying
\eqref{eq:=00005BExpect=00005D}-\eqref{eq:39} is illustrated in
algorithm-1.

\begin{algorithm}
\DontPrintSemicolon
\SetAlgoNlRelativeSize{0}
\SetNlSty{}{}{}

\KwIn{Initial phase shifts $\Theta$, Path-loss $L_{\mathrm{RU_k}}$, Other system parameters.}
\KwOut{Optimal energy efficiency $\eta^*$ and optimal phase shifts $\Theta^*$.}

Initialize $\Theta$ with random values in range $[0,2\pi]$ for each RIS element.\;
Initialize moment estimates $M_t$ and $V_t$ to zero.\;
Set iteration count $t = 0$.\;
Set small value $\epsilon = 1\times10^{-8}$.\;

\While{not converged}{
    Compute gradient $g_t$ using Monte Carlo approximation: $g_t = \frac{1}{M}\sum_{m=1}^{M} \frac{\partial\eta(\Theta, L_{\mathrm{RU_k}}^{(m)})}{\partial\phi_{kn}}$ \;
    Update moment estimates: $M_t = \beta_1 M_{t-1} + (1-\beta_1) g_t$, $V_t = \beta_2 V_{t-1} + (1-\beta_2) g_t^2$ \;
    Correct bias in moment estimates: $\hat{M_t} = \frac{M_t}{1-\beta_1^t}$, $\hat{V_t} = \frac{V_t}{1-\beta_2^t}$ \;
    Update phase shifts: $\phi_{kn}^{(t+1)} = \phi_{kn}^{(t)} + \alpha \frac{\hat{M_t}}{\sqrt{\hat{V_t}}+\epsilon}$ \;
    Clip updated phase shifts: $\phi_{kn}^{(t+1)} = \text{clip}(\phi_{kn}^{(t+1)},0,2\pi)$ \;
    Increment iteration count $t = t + 1$.\;
}

Compute $\eta^* = \eta(\Theta^*,L_{\mathrm{RU_k}})$ using the optimized phase shifts $\Theta^*$.\;

\caption{Optimization of Energy Efficiency using Adam Algorithm}
\end{algorithm}

\begin{figure}
\centering{}\includegraphics[width=3.5in]{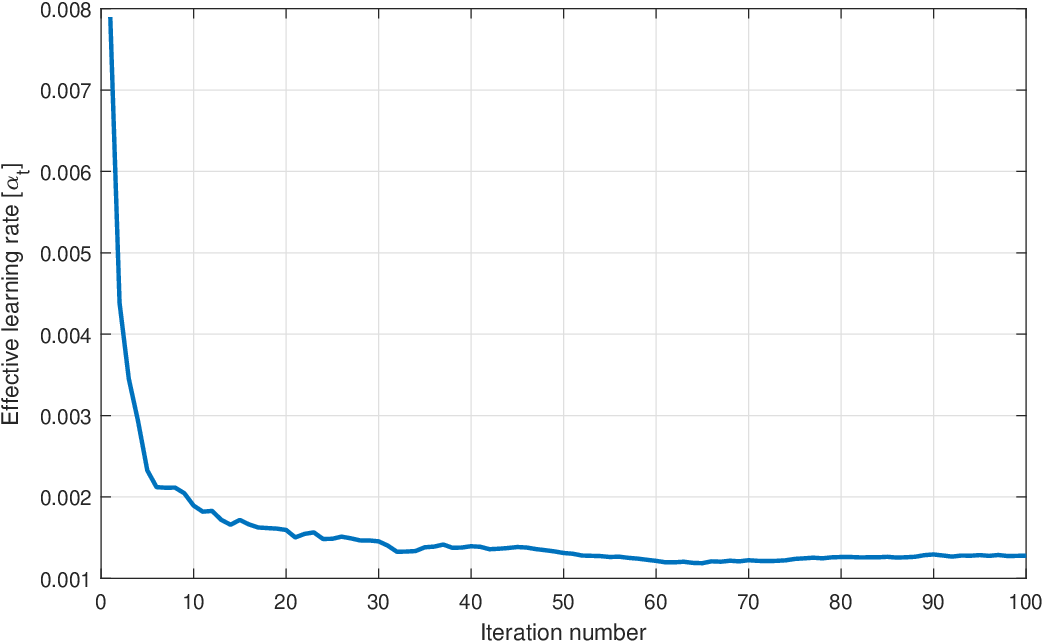}\caption{Evolution of learning rate during the optimization process.\label{fig: 6}}
\end{figure}

The graphical representation in Fig. \ref{fig: 6} depicts how the
learning rate, $\alpha_{t}$, evolves during the optimization process
using the Adam algorithm. The x-axis, labeled \textquotedbl Iteration\textquotedbl ,
counts the number of updates made to the $\Theta$ aiming to enhance
energy efficiency. The y-axis, titled $\alpha_{t}$, shows the adjusted
learning rate at each step. Initially, the learning rate is set higher,
enabling the algorithm to swiftly explore the solution space and make
significant changes to $\Theta$. As the process continues, the learning
rate gradually decreases, allowing for more precise, smaller adjustments
to $\Theta$. Such modulation ensures that the algorithm zeroes in
on the optimal solution without undue fluctuations or overshooting.
The declining trend of the learning rate is a characteristic feature
of the Adam algorithm. It ensures that, at the outset, the algorithm
explores solutions rapidly. However, as it nears the optimal solution,
it adopts a more cautious approach, ensuring a steady and reliable
convergence.

\begin{figure}
\centering{}\includegraphics[width=3.5in]{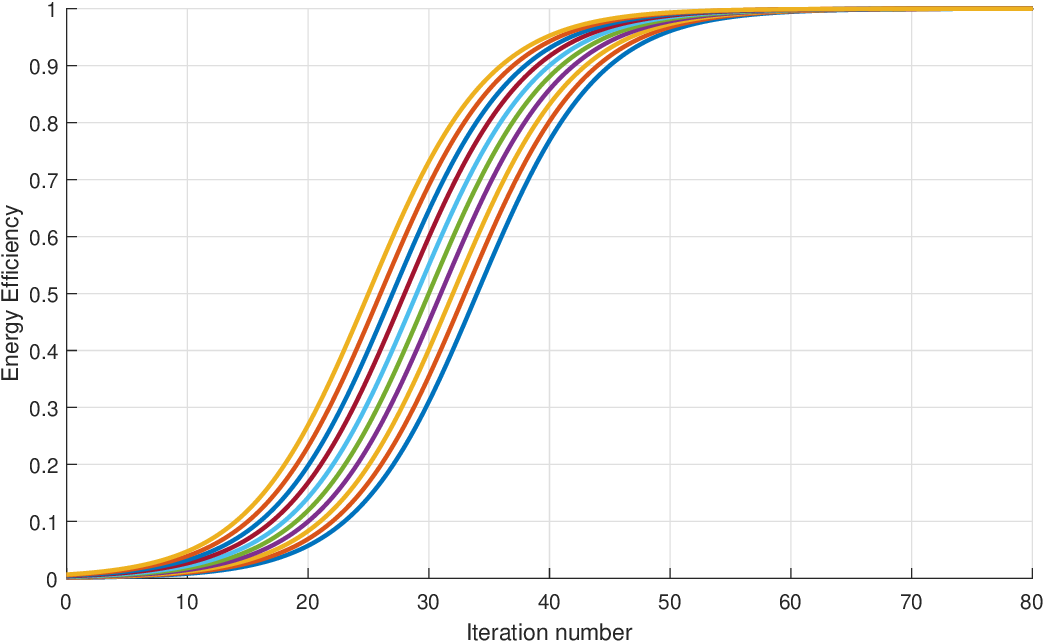}\caption{Convergence behaviour of Algorithm-1 with 10 different initial solutions\label{fig: 7 }}
\end{figure}

Figure \ref{fig: 7 } illustrates the progression of energy efficiency
over iterative optimization using the Adam algorithm, considering
various initial phase shift setups for the RISs. Each trajectory in
the plot represents a distinct initial configuration of phase shifts,
$\Theta$, for the RIS elements. The y-axis quantifies the energy
efficiency, while the x-axis, labeled \textquotedbl Iteration\textquotedbl ,
indicates the number of applications of the Adam algorithm to refine
and update the phase shifts, $\Theta$, with the objective of optimizing
energy efficiency. A close examination of Fig. \ref{fig: 7 } reveals
that, irrespective of the initial configurations, all trajectories
converge to a similar energy efficiency value. This behaviour underscores
the robustness of the Adam optimization algorithm, which dynamically
adjusts its learning rates based on the gradient's first and second
moment estimates. The algorithm's adaptability ensures that even in
the presence of stochastic variables like shadowing effect, it can
navigate the optimization landscape effectively and converge to an
optimal or near-optimal solution. The convergence of all trajectories
to a similar value also suggests that the system's energy efficiency
has a global optimum that is reachable from various initial configurations.
This is a promising observation, indicating that the system's performance
is not overly sensitive to initial conditions, and the optimization
process is effective in enhancing energy efficiency across different
scenarios.

\begin{figure}
\centering{}\includegraphics[width=3.5in]{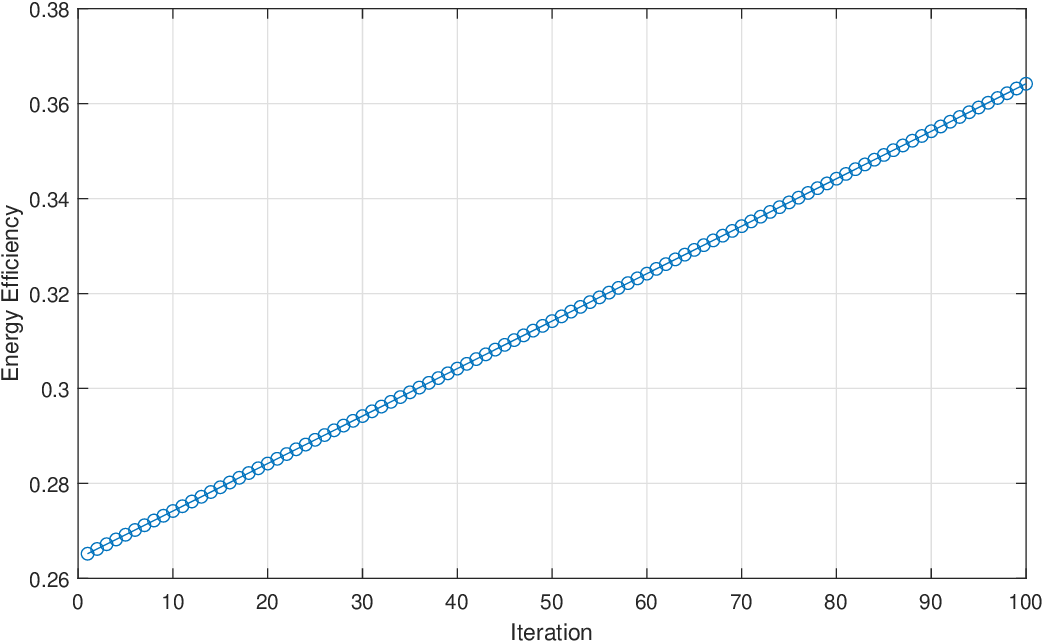}\caption{Tracking energy efficiency progression.\label{fig: 8}}
\end{figure}

To provide a clearer understanding of the optimization process and
its impact on system performance, we turn our attention to Fig. \ref{fig: 8}.
This figure visually depicts the evolution of energy efficiency across
iterations. The plotted line's rising and subsequent stabilizing pattern
suggests that the algorithm is effectively refining the phase shifts,
leading towards an optimal or near-optimal solution. This behavior
underscores the algorithm's capability to navigate the solution space
and make beneficial adjustments, ultimately achieving a desirable
energy efficiency outcome.

\begin{figure}
\centering{}\includegraphics[width=3.5in]{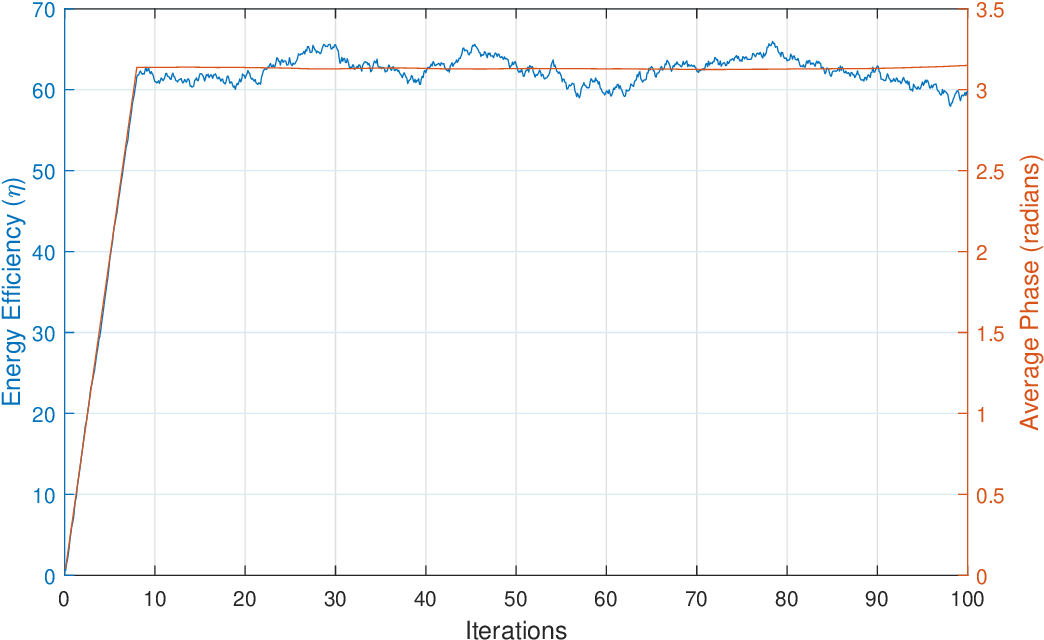}\caption{the energy efficiency, $\mathbb{E}[\eta]$, vs. RIS-User and phase
Shift\label{fig: 9}}
\end{figure}

\begin{figure}
\centering{}\includegraphics[width=3.5in]{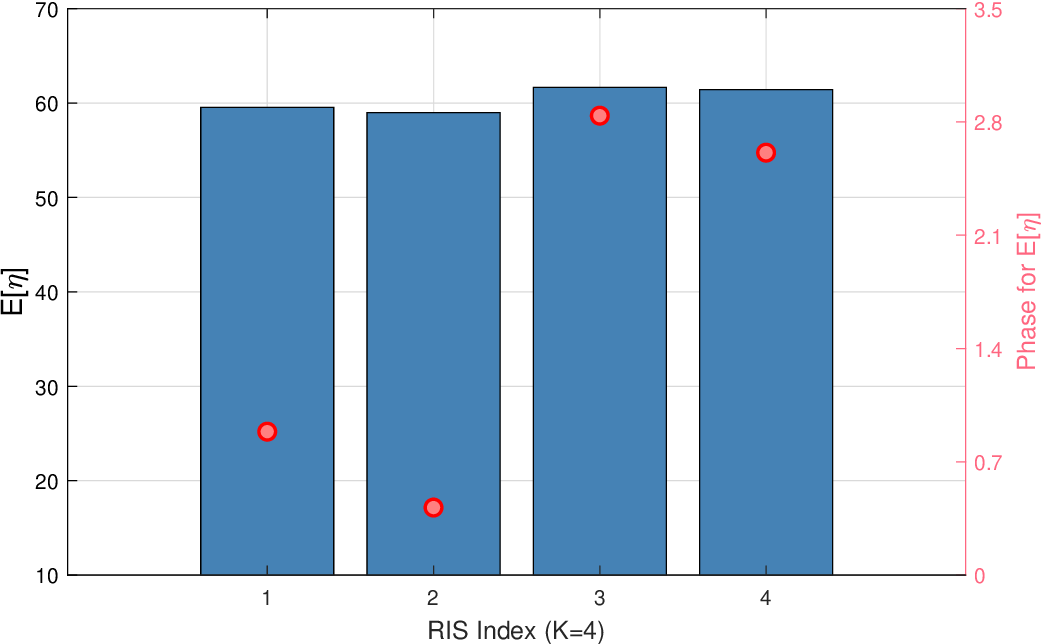}\caption{Maximum $\mathbb{E}[\eta]$ for Each RIS based on phase values\label{fig: 10}}
\end{figure}

To elucidate the modulation of phase configurations of RIS elements
by the Adam algorithm for the maximization of energy efficiency, Fig.
\ref{fig: 9} presents a dynamic illustration of the optimization
process over time. In this figure, the curve representing energy efficiency
initially ascends and subsequently stabilizes. This pattern indicates
that the algorithm is progressively optimizing the phase configurations
to enhance energy efficiency, eventually converging to an optimal
or near-optimal solution. In parallel, the phase curve provides insights
into the average adjustments made to the RIS elements over time. Notably,
this curve reflects the trend of the energy efficiency curve, highlighting
that the algorithm persistently modifies the configurations until
a stable phase setting is achieved. While Fig. \ref{fig: 9} unveils
the dynamics of the optimization process, Fig. \ref{fig: 10} delineates
the final outcomes, detailing the maximal energy efficiency realized
by each RIS at specific phase values. This figure integrates a bar
chart that exhibits the maximal $\mathbb{E}[\eta]$ for every RIS,
calculated from normalized phase values. The height of each bar corresponds
to the energy efficiency attained by an individual RIS subsequent
to phase optimization. Superimposed on this bar chart is a line graph
with markers, each of which denotes the optimal phase shift at which
the associated RIS reaches its peak energy efficiency. Annotations
on both the bars and markers provide exact values, facilitating a
comprehensive understanding of the results.

\section{Conclusion\label{Conclusion} }

This study presents two innovative approaches designed to enhance
energy efficiency in satellite-to-ground communication systems, utilizing
multiple Reflective Intelligent Surfaces (RISs). The central objective
of these methods is to maxmize the overall energy efficiency of the
system, and this is achieved by tacking two distinct challenges. In
the first approach, denoted as IE, the emphasis is on amplifying power
reception by fine-tuning the phase shift of each RIS. Subsequently,
Selective Diversity is used to identify the RIS delivering the maximum
power. The next challenge revolves around minimizing power consumption.
This challenge is articulated as a binary linear programming problem
and tackled using the BPSO technique. The IE approach presumes an
ideal environment where signals propagate without any path loss. This
ideal scenario is used as a foundational reference for theoretical
evaluations, shedding light on the peak capabilities of the system.
In contrast, the second approach, termed NIE, is designed for scenarios
where signal transmission is subject to path loss. Within this framework,
the Adam algorithm is utilized to optimize energy efficiency. This
non-ideal setting offers a pragmatic evaluation of the system's capabilities
under conventional operational conditions. The efficacy of the proposed
framework is assessed under these two paradigms. Both scenarios underscore
the potential energy savings of the satellite-RIS system. The ideal
setting elucidates the system's pinnacle performance under optimal
conditions, while the real-world scenario underscores the practical
challenges and requisite adaptations. Importantly, the framework integrates
probabilistic considerations to accommodate external environmental
variables, representing a substantial advancement in optimizing energy
efficiency for RIS-augmented satellite communication.

\bibliographystyle{IEEEtran}
\addcontentsline{toc}{section}{\refname}\bibliography{RMIT}

\end{document}